\documentclass[pra,twocolumn]{revtex4}
\usepackage{amsfonts,amsmath,amssymb,float}
\usepackage{bm,dsfont}
\usepackage{color}
\usepackage{bbm}
\usepackage{longtable}
\usepackage[dvips]{epsfig}
\usepackage{amsmath,amssymb,lscape,float}
\usepackage{hyperref}
\usepackage{listings}

\usepackage{graphicx}
\usepackage{epstopdf}
\usepackage{dcolumn}
\usepackage{bm}
\usepackage{amsfonts,amsmath,amssymb
}

\usepackage{bm,dsfont}
\usepackage{color}
\usepackage{bbm}
\usepackage{longtable}
\usepackage[dvips]{epsfig}
\usepackage{hyperref}
\usepackage{listings}
\usepackage{todonotes}

\newcommand{\bra}[1]{\langle #1|}

\newcommand{\ket}[1]{|#1 \rangle}

\begin{document}
\title{Interconversion of $W$ and Greenberger-Horne-Zeilinger states for 
Ising-coupled qubits with transverse global control }
\author{Vladimir M. Stojanovi\'c and Julian K. Nauth}
\affiliation{Institut f\"{u}r Angewandte Physik, Technical
University of Darmstadt, D-64289 Darmstadt, Germany}
\date{\today}
\begin{abstract}
Interconversions of $W$ and Greenberger-Horne-Zeilinger states in various physical systems are lately attracting 
considerable attention. We address this problem in the fairly general physical setting of qubit arrays with long-ranged
(all-to-all) Ising-type qubit-qubit interaction, which are simultaneously acted upon by transverse Zeeman-type 
global control fields. Motivated in part by a recent Lie-algebraic result that implies state-to-state controllability 
of such a system for an arbitrary pair of states that are invariant with respect to qubit permutations, 
we present a detailed investigation of the state-interconversion problem in the three-qubit case. The envisioned 
interconversion protocol has the form of a pulse sequence that consists of two instantaneous (delta-shaped) control 
pulses, each of them corresponding to a global qubit rotation, and an Ising-interaction pulse of finite duration 
between them. Its construction relies heavily on the use of the (four-dimensional) permutation-invariant subspace 
(symmetric sector) of the three-qubit Hilbert space. In order to demonstrate the viability of the proposed 
state-interconversion scheme, we provide a detailed analysis of the robustness of the underlying pulse sequence 
to systematic errors, i.e. deviations from the optimal values of its five characteristic parameters.
\end{abstract}
	
\maketitle
\section{Introduction}
Regardless of their concrete physical realization, maximally-entangled multiqubit states are of utmost importance for 
quantum-information processing (QIP)~\cite{NielsenChuangBook}. Two prominent classes of such states, 
which cannot be transformed into each other through local operations and classical communication~\cite{Nielsen:99}, are 
$W$~\cite{Duer+:00} and Greenberger-Horne-Zeilinger (GHZ)~\cite{Greenberger+Horne+Zeilinger:89} states. In particular, in 
the three-qubit case $W$ and GHZ are the only two subclasses of states with genuine tripartite entanglement~\cite{Zhu+:22}. 
Both classes have proven useful in diverse QIP contexts~\cite{Joo+:03,Zhu+:15,Lipinska+:18,Sun+:20,Omkar+:22}, which was 
the primary motivation behind a large number of proposals for the efficient preparation of $W$~\cite{Peng+:0910,Dogra+:14,
Li+Song:15,Kang+:16,Kang+SciRep:16,Fang+:19,StojanovicPRL:20,Peng+:21,StojanovicPRA:21,Zheng++:22,Zhang+:22} and GHZ 
states~\cite{Coelho+:09,Wang+:10,Dogra+:14,Song+:17,Erhard+:18,Macri+:18,Zheng+:19,Pachniak+Malinovskaya:21,
Nogueira+:21,Qiao+:22} in various physical systems. 

In view of the completely different characters of entanglement in $W$ and GHZ states~\cite{Duer+:00,HorodeckiRMP:09}, the 
interconversion between those states in different physical platforms represents an interesting, increasingly relevant problem of 
quantum-state engineering. The earliest attempt in the context of such an interconversion pertained to a photonic system~\cite{Walther+:05}. 
This initial study, which was probabilistic in nature, was followed by another photon-related work~\cite{Cui+:16} and an investigation
of such interconversions in a spin system~\cite{Kang+:19}. In the realm of atomic systems, irreversible conversions of $W$- into
GHZ states were first proposed~\cite{Song+:13,Wang+:16}. More recently, the deterministic interconversion between the two states
in a system of three Rydberg-atom-based qubits~\cite{Morgado+Whitlock:21,ShiREVIEW:22} subject to four external laser pulses was 
extensively studied~\cite{Zheng+:20,Haase+:21,Haase++:22,Nauth+Stojanovic:22}.
	
In this paper, the interconversion of $W$ and GHZ states problem is addressed for an array of qubits coupled through Ising-type 
($ZZ$) interaction, being also subject to two Zeeman-like global control fields in the transverse ($x$- and $y$) directions. 
The Ising-type coupling between qubits is of practical importance as it enables the realization of the controlled-$Z$ gate
(also known as the controlled phase-shift gate~\cite{Jones:03}). Namely, the Ising-coupling gate and controlled-$Z$ are
related by single-qubit $z$ rotations and a global phase shift~\cite{Jones:03}. At the same time, the controlled-$Z$ gate
differs from controlled-NOT (CNOT) only by two Hadamard gates applied to the target qubit of the CNOT gate~\cite{NielsenChuangBook}. 

It is worthwhile pointing out that, generally speaking, global-control schemes for qubit arrays constitute a promising pathway 
towards scalable QC. Apart from obviating the need for local qubit addressing, which in some physical platforms for QC is 
unfeasible, another well-known advantage of such schemes stems from the fact that a continuous-wave global field can efficiently 
decouple qubits from the background noise~\cite{Jones+:18}. 

The motivation behind the present work is twofold. Firstly, a qubit array with long-range Ising-type qubit-qubit interactions 
can be realized in various physical platforms for QC, from nuclear-magnetic-resonance (NMR) systems~\cite{Vandersypen+Chuang:05,
Viola+:01,Fortunato+:05} to ensembles of neutral atoms in Rydberg states~\cite{Morgado+Whitlock:21}; therefore, an efficient 
solution of the  $W$-to-GHZ state-conversion problem may facilitate 
the realization of various QIP protocols in those systems. Secondly, a recent result in the realm of Lie-algebraic controllability 
implies that such an array of Ising-coupled qubits, which is subject to global control fields in the two transverse directions, is
indeed state-to-state controllable provided that the two relevant (initial and final) states are invariant under an arbitrary
permutation of qubits~\cite{Albertini+DAlessandro:18}; moreover, it is important to note that both $W$ states and their GHZ 
counterparts are permutationally invariant for an arbitrary number of qubits.

While the aforementioned Lie-algebraic result~\cite{Albertini+DAlessandro:18} guarantees the existence of a quantum-control protocol 
for converting a $W$ state into its GHZ counterpart for Ising-coupled qubits with global transverse control, a solution of the last 
problem for a three-qubit system is presented in this paper. The envisioned state-conversion protocol is based on an NMR-type pulse 
sequence that consists of two instantaneous (delta-shaped) global control pulses and an Ising-interaction pulse of finite duration 
between them. The construction of this pulse sequence, as well as its robustness against errors in the relevant parameters (e.g. small 
variations of the duration of Ising-interaction pulses and global-rotation angles corresponding to the transverse control fields), 
are discussed in detail in what follows. It is worthwhile mentioning that pulse sequences of this kind have as yet been utilized 
in multiple physical contexts of interest for QIP~\cite{Jones:03,Hill:07,Geller+:10,Ghosh+Geller:10,TanamotoQECC}. For example, they 
were proposed by one of us and collaborators for applications in measurement-based quantum computing~\cite{Raussendorf+Briegel:01}, 
more precisely for preserving cluster states~\cite{Tanamoto+:12}, as well as for dynamically generating code words of various 
quantum error-correction codes~\cite{Tanamoto+:13}.

The remainder of the present paper is organized as follows. In Sec.~\ref{SystemProblem}, the system under consideration and 
the state-conversion problem to be addressed in the following are introduced, along with the notation to be used throughout 
the paper. Section~\ref{SymmSector} is devoted to the symmetry-related aspects of the problem at hand, more precisely its 
invariance under an arbitrary permutation of qubits and the ensuing concept of the symmetric sector of the three-qubit Hilbert 
space. In addition, one familiar (symmetry-adapted) basis of the latter subspace is introduced. In Sec.~\ref{WtoGHZsequence} 
the construction of an NMR-type pulse sequence, which represents one solution of the state-conversion problem in the three-qubit
case, is discussed in detail. The principal results for the idealized pulse sequence behind the $W$-to-GHZ state conversion, 
as well as its robustness to errors in its characteristic parameters, are presented in Sec.~\ref{StateConvResDisc} Finally, 
the paper is summarized -- along with underscoring its main conclusions and possible generalizations -- in Sec.~\ref{SummConcl}.

\section{System and $W$-to-GHZ conversion problem} \label{SystemProblem}
The system under consideration is a qubit array with long-range Ising-type coupling with 
strength $J$, subject to global Zeeman-type control fields $h_x(t)$ and $h_y(t)$ in the $x$- and 
$y$ directions, respectively. The total Hamiltonian of the system $H(t)=H_{ZZ}+H_C(t)$ consists of
the drift (Ising-interaction) part $H_{ZZ}$ and the global-control part $H_C(t)$. It can 
succinctly be written as
\begin{equation} \label{TotalHamiltonian}
H(t)=H_{ZZ} + h_x(t)\mathcal{X}+h_y(t)\mathcal{Y}\:.
\end{equation}
Here $H_{ZZ}$, $\mathcal{X}$, and $\mathcal{Y}$ are given by
\begin{eqnarray}
H_{ZZ} &=& J\sum_{1\le n<n'\le N}\:Z_n Z_{n'} \:, \label{LRIsingInt} \\
\mathcal{X} &=& \sum_{n=1}^{N}\:X_n \quad,\quad \mathcal{Y} = 
\sum_{n=1}^{N}\:Y_n\:, \label{ControlOperators}
\end{eqnarray}
where $X_n$, $Y_n$, and $Z_n$ are the Pauli operators of qubit $n$ ($n=1,\ldots,N$):
\begin{eqnarray} \label{defineXYZ_n}
X_n &=& \mathbbm{1}\otimes\ldots\otimes\mathbbm{1}\otimes 
\underbrace{X}_n\otimes\mathbbm{1}\otimes\ldots\otimes\mathbbm{1}\: ,\notag\\
Y_n &=& \mathbbm{1}\otimes\ldots\otimes\mathbbm{1}\otimes 
\underbrace{Y}_n\otimes\mathbbm{1}\otimes\ldots\otimes\mathbbm{1}\: ,\\
Z_n &=& \mathbbm{1}\otimes\ldots\otimes\mathbbm{1}\otimes 
\underbrace{Z}_n\otimes\mathbbm{1}\otimes\ldots\otimes\mathbbm{1} \notag \:.
\end{eqnarray}

It is pertinent to comment on the controllability~\cite{D'AlessandroBook} aspects of systems described by the
Hamiltonian of Eq.~\eqref{TotalHamiltonian}. In this context, it is useful to first point out that for complete 
operator controllability (implying the ability to realize an arbitrary unitary transformation on the Hilbert 
space of the underlying system, i.e. universal quantum computation) of a qubit array with Ising-type interaction
it is required to have two mutually noncommuting (local) controls acting on each qubit in the array~\cite{Wang++:16}. 
In fact, it is only for qubit arrays with Heisenberg-type interaction (isotropic, $XXZ$-, or $XYZ$-type) that 
a significantly reduced degree of control -- namely, two noncommuting controls acting on a single qubit in the 
array -- guarantees complete controllability~\cite{Heule++:10,Wang++:16}. Thus, a system of $N$ qubits that are 
coupled through Ising-type interaction and subject to global Zeeman-like control fields in the $x$- and $y$ directions 
[cf. Eqs.~\eqref{TotalHamiltonian}-\eqref{ControlOperators} above], is in general not completely operator controllable; 
in other words, its dynamical Lie algebra~\cite{D'AlessandroBook} $\mathcal{L}_{d}=\textrm{span}\{H_{ZZ},
\mathcal{X},\mathcal{Y}\}$ is not isomorphic with $u(2^N)$ or $su(2^N)$, but with their proper Lie subalgebra. 

Despite the lack of complete controllability, it has recently been demonstrated that a system described by the Hamiltonian 
in Eq.~\eqref{TotalHamiltonian}, which is manifestly symmetric with respect to an arbitrary permutation of qubits (i.e. 
spin-$1/2$ subsystems), is controllable provided one restricts oneself to unitary evolutions that preserve this permutation 
invariance~\cite{Albertini+DAlessandro:18}. An immediate implication of this last result is that such a system is 
state-to-state controllable for any pair of states that are themselves invariant with respect to qubit permutations. 
This is equivalent to the statement that the time-dependence of control fields $h_x(t)$ and $h_y(t)$ in Eq.~\eqref{TotalHamiltonian} 
can be found such that one can reach any permutationally invariant final state in a finite time starting from an arbitrary 
permutationally invariant state at $t=0$. As usual for Lie-algebraic controllability theorems~\cite{D'AlessandroBook}, which 
have the character of existence theorems, the actual time-dependence of these control fields that enables a controlled 
dynamical evolution of the system from a given initial- to a desired final state has to be determined in each particular 
case~\cite{Zhang+Whaley:05}.

In what follows, we design protocols for the deterministic interconversion of $W$ and GHZ state in a three-qubit 
system ($N=3$). The general expressions of Eqs.~\eqref{LRIsingInt} and \eqref{ControlOperators} in that case 
reduce to
\begin{eqnarray} \label{threeQubitZZXY}
H_{ZZ} &=& J(Z_1 Z_2+Z_2 Z_3+Z_1 Z_3) \nonumber \:,\\
\mathcal{X} &=& X_1+X_2+X_3 \:,\\
\mathcal{Y} &=& Y_1+Y_2+Y_3 \nonumber \:,
\end{eqnarray}
where -- in line with the general definition of $X_n,Y_n,Z_n$ [cf. Eq.~\eqref{defineXYZ_n}] -- the operators 
$X_1,X_2\ldots,Z_3$ are represented in the standard computational basis by eight-dimensional matrices.

Because $W$ and GHZ states both have the property of being permutation-invariant (for an arbitrary number of qubits),
our treatment of the state-conversion problem for a three-qubit system will rely heavily on this last property of the 
initial and final states. More precisely, in the following a protocol is sought after that allows the conversion of an 
initial $W$ state into a GHZ state; the inverse state-conversion process -- converting an initial GHZ state into its 
$W$-state counterpart -- is analyzed in an analogous fashion. In other words, the state $\ket{\psi(t)}$ of our three-qubit 
system should satisfy the conditions 
\begin{eqnarray}\label{InitFinalStates}
\ket{\psi(t=0)} &=& \ket{W_3}=\frac{1}{\sqrt{3}}\left(|100\rangle+|010\rangle+|001\rangle\right)\:, \label{eq:psit}\\
\ket{\psi(t=T)} &=&\ket{\textrm{GHZ}_3(\varphi)}=\frac{1}{\sqrt{2}}\left(\ket{000} + e^{i\varphi}\ket{111}
\right)  \nonumber \:,
\end{eqnarray}
where $\varphi\in[0,2\pi)$ and $T$ is the state-conversion time. 

The $W$-to-GHZ state conversion will be achieved here using an NMR-type pulse sequence. Such sequences consist
of a certain number of instantaneous (delta-shaped) control pulses and Ising-interaction pulses in between the 
control pulses. In the following, we set $\hbar=1$, hence all the relevant timescales in the problem at hand 
will be expressed in units of the inverse Ising-coupling strength $J^{-1}$.

\section{Symmetric sector and its basis} \label{SymmSector}
In what follows, we describe the problem under consideration by exploiting its permutation-symmetric
character. To this end, we first introduce the concept of the symmetric sector of the three-qubit Hilbert 
space and define one specific basis of this sector that facilitates the solution of the state-conversion 
problem at hand. 

In a variety of problems in quantum control and quantum-state engineering it is beneficial to consider pure states 
that are invariant with respect to permutations of qubits~\cite{Zanardi:99,Ribeiro+Mosseri:11,Burchardt+:21,
Chryssomalakos+:21,Hebenstreit+:22,Lyons+:22}. In this context, we can distinguish situations where the relevant 
states are those invariant under an arbitrary permutation -- i.e. the full symmetric group $S_n$, where $n$ is 
the number of qubits~\cite{Hebenstreit+:22} -- and those where the relevant states are invariant with respect
to specific nontrivial subgroups of $S_n$~\cite{Lyons+:22}.

In the state-conversion problem at hand, we focus on the subset of all the unitaries on the Hilbert space 
$\mathcal{H}\equiv(\mathbbm{C}^2)^{\otimes 3}$ of the three-qubit system under consideration that are invariant 
under an arbitrary qubit permutation, i.e. the permutation group $S_3$. The relevant Lie subgroup of $U(8)$ 
is denoted by $U^{\textrm{S}_3}(8)$ and has dimension equal to $20$~\cite{Albertini+DAlessandro:18}. Its 
corresponding Lie algebra $u^{\textrm{S}_3}(8)$ is spanned by the operators $i\varPi(\sigma_1\otimes\sigma_2
\otimes\sigma_3)$, where $\varPi=(3!)^{-1}\:\sum_{P\in S_3}\:P$ and $\sigma_n$ ($n=1,2,3$) is either the 
single-qubit identity operator $\mathbbm{1}_{2}$ or one of the Pauli operators $X$, $Y$, $Z$. 

Under the action of the Lie algebra $u^{\textrm{S}_3}(8)$ the $8$-dimensional Hilbert space $\mathcal{H}$ 
splits into three invariant subspaces that correspond to irreducible representations of $su(2)$. Two of
those subspaces have dimension $2$, while the third one has dimension $4$ and is uniquely determined. The 
latter is usually referred to as the {\em symmetric sector}~\cite{Ribeiro+Mosseri:11}, because it comprises 
the states that do not change under an arbitrary permutation of qubits. One orthonormal, symmetry-adapted 
basis of the symmetric sector is given by the states $\{|\zeta_a\rangle|\:a=0,\ldots,3 \}$, where
\begin{eqnarray} \label{SSbasis}
|\zeta_0\rangle = |000\rangle \:,\quad |\zeta_1\rangle=\frac{1}{\sqrt{3}}\:
(|100\rangle+|010\rangle+|001\rangle) \:,\\
|\zeta_2\rangle=\frac{1}{\sqrt{3}}\:(|110\rangle+|101\rangle+|011\rangle) 
\:,\quad |\zeta_3\rangle = |111\rangle  \:, \nonumber
\end{eqnarray}
and the subscript $a$ in $|\zeta_a\rangle$ coincides with the Hamming weight of the corresponding 
bit string (i.e. the number of occurrences of $1$ in that bit string)~\cite{Kim:18}. It is obvious 
that $|\zeta_1\rangle\equiv|W_3\rangle$ is the $W$ state itself, while $|\zeta_2\rangle$ corresponds
to the two-excitation Dicke state.

In the following, we consider the state-conversion problem within the symmetric sector using the basis 
defined in Eq.~\eqref{SSbasis}. To begin with, we map the four basis states onto column vectors according 
to
\begin{eqnarray} \label{BasisColumnVecs}
\ket{\zeta_0}&\mapsto&
\left(\begin{array}{c}
1\\
0\\
0\\
0
\end{array}\right),~~\ket{\zeta_1} ~\mapsto~
\left(\begin{array}{c}
0\\
1\\
0\\
0
\end{array}\right),\nonumber\\
\ket{\zeta_2} &\mapsto&
\left(\begin{array}{c}
0\\
0\\
1\\
0
\end{array}\right),~~\ket{\zeta_3}~\mapsto~
\left(\begin{array}{c}
0\\
0\\
0\\
1
\end{array}\right).
\label{BasisinSS}
\end{eqnarray}
We also can straightforwardly represent the initial and target states of our envisioned state conversion 
[cf. Eq.~\eqref{eq:psit}] in this same basis. While $|\zeta_1\rangle\equiv|W_3\rangle$, the GHZ state is
given by
\begin{equation}
\ket{\textrm{GHZ}_3(\varphi)} \mapsto
\frac{1}{\sqrt{2}}
\left(\begin{array}{c}
1\\
0\\
0\\
e^{i\varphi}
\end{array}\right)\:. \label{WandGHZinSS}
\end{equation}

For the sake of completeness, it is worthwhile mentioning that a generalized Schmidt decomposition 
allowed a classification of pure three-qubit states~\cite{Acin+:00}. More specifically yet, in Ref.~\cite{Acin+:00} 
it was demonstrated that five independent nonzero real parameters are needed to describe the entire three-qubit 
state space under local operations; in other words, a generic pure three-qubit state is equivalent under local 
unitary transformations to a canonical state described by these five parameters. It was shown that there exist, 
in fact, three inequivalent sets of five local basis product states, where each of these three sets contains the 
states $|000\rangle,\:|100\rangle$, and $|111\rangle$. One of those sets, given by $\{|000\rangle,\:|001\rangle,
\:|010\rangle,\:|100\rangle,|111\rangle\}$, is symmetric with respect to permutations of qubits (parties) and yields 
three-qubit $W$ and GHZ states as linear combinations of its elements. It is also worthwhile pointing out that 
an experimental scheme for creating a generic pure three-qubit state in NMR -- in line with the classification 
in Ref.~\cite{Acin+:00} -- was proposed in the past~\cite{Dogra+:14}.

\section{$W$-to-GHZ state conversion using a pulse sequence} \label{WtoGHZsequence}
We aim to find a solution of the $W$-to-GHZ state conversion problem [cf. Eq.~\eqref{InitFinalStates}]
for an arbitrary value of $\varphi$. As indicated above, the two states of interest are invariant
with respect to an arbitrary permutation of qubits. Thus, the problem can be reduced to the symmetric 
sector and its basis given in Eq.~\eqref{SSbasis} above.

In the following, we first describe the layout of the envisioned pulse sequence for implementing $W$-to-GHZ 
state conversion (Sec.~\ref{PulseSequence}), followed by the derivation of the time-evolution operators 
corresponding to different parts of this pulse sequence (Sec.~\ref{TimeEvolOps}).
\subsection{Form of the pulse sequence} \label{PulseSequence}
We seek a solution to the $W$-to-GHZ state conversion problem in the form of an NMR-type pulse sequence 
that consists of two instantaneous control pulses -- at times $t=0$ and $t=T$ (i.e. with a time delay 
$T$ between them) -- and an Ising-interaction pulse with duration $T$ between these control pulses (for
a pictorial illustration, see Fig.~\ref{fig:PulseSequence} below). The corresponding transverse (global)
control field $\boldsymbol{h}(t)\equiv[h_x(t), h_y(t),0]^{\textrm{T}}$ can be written as
\begin{equation}\label{eq:h(t)_def}
\boldsymbol{h}(t)=\boldsymbol{\alpha}_1\delta(t)+\boldsymbol{\alpha}_2\delta(t-T) \:,
\end{equation}
where the two delta functions capture the instantaneous character of the two control pulses and the vectors 
$\boldsymbol{\alpha}_1$ and $\boldsymbol{\alpha}_2$ point in arbitrary directions in the $x$-$y$ plane; the 
corresponding directions are specified by their polar angles $\phi_1$ and $\phi_2$, respectively, where $\phi_1,
\phi_2\in [0,2\pi)$. 

Before embarking on the derivation of the respective time-evolution operators that correspond to different 
parts of the envisioned pulse sequence (cf. Fig.~\ref{fig:PulseSequence}), it is pertinent to comment on the 
feasibility of realizing such pulse sequences in various physical platforms for QC. Firstly, the assumption 
of instantaneous control pulses is well justified whenever the control fields used are much stronger than 
the coupling between qubits; this requirement is, for example, satisfied for typical control magnetic 
fields used in the NMR realm~\cite{Vandersypen+Chuang:05}, as well as for typical control fields in 
superconducting-qubit-~\cite{Geller+:10} and neutral-atom systems~\cite{Morgado+Whitlock:21}. Secondly, the 
fact that the envisioned pulse sequence entails single-qubit rotations about two different axes in the $x$-$y$ 
plane is feasible in practice. Namely, modification of the rotation axis of a single-qubit drive represents 
a rather straightforward operation in currently used platforms for QC, such as neutral atoms~\cite{Xia+:15}, 
superconducting qubits~\cite{McKay+:17}, and trapped ions~\cite{Debnath+:16}; such an operation does not
involve much of an additional experimental complexity or overhead.

\begin{figure}[t!]
\includegraphics[width=0.975\linewidth]{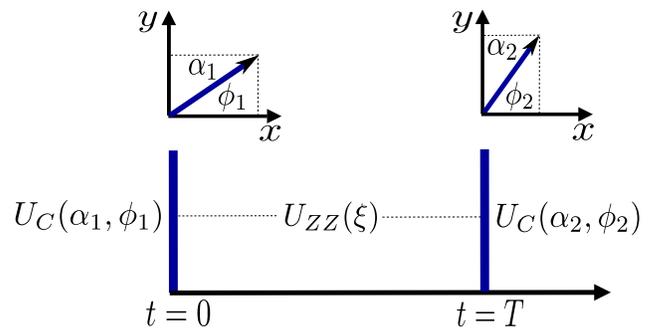}
\caption{\label{fig:PulseSequence}(Color online)Pictorial illustration of the pulse sequence for
realizing the $W$-to-GHZ state conversion, which consists of two instantaneous control pulses and 
an Ising-interaction pulse of finite duration $T$ between them. The first (second) control pulse is
characterized by a vector in the $x$-$y$ plane, with the magnitude $\alpha_1$ ($\alpha_2$) and polar 
angle $\phi_1$ ($\phi_2$). Here $U_C(\alpha_1,\phi_1)$ and $U_C(\alpha_2,\phi_2)$ are the time-evolution
operators corresponding to the control pulses at $t=0$ and $t=T$, respectively; $U_{ZZ}(\xi)$ 
corresponds to the interaction pulse, with $\xi\equiv JT$ being its dimensionless duration.}
\end{figure}

It is important to stress that the global character of single-qubit rotations in the envisioned pulse 
sequence is, in fact, a necessity in many QC platforms under current investigation. An example is furnished by a typical 
setup for neutral-atom QC, both in cases where the role of two logical states of a qubit is played by two hyperfine 
states ($gg$ qubits) and in cases where a ground state and a high-lying Rydberg state play this role ($gr$ qubits).
In such neutral-atom systems one typically makes use of a global microwave field -- which in the case of $gg$ qubits 
has the physical nature of magnetic dipole coupling -- to carry out a rotation about an arbitrary axis in the $x-y$ plane 
on every qubit~\cite{Graham+:19}; at the same time, the rotation axis can be chosen by modifying the phase of the microwave 
field. Importantly, this rotation gate ought to be global in nature because the distance between qubits in such systems
(typically a few micrometers) is far smaller than the wavelength of the microwave field ($1\:\textrm{mm}\lesssim \lambda
\lesssim 1\:\textrm{m}$); as a result, each qubit undergoes the same rotation. 

While here we aim for an analog implementation of the envisioned pulse sequence, it should be stressed that 
such a pulse sequence is also amenable to an efficient digital realization in various QC platforms. For example, in the 
neutral-atom platform a rotation over an arbitrary axis in the $x-y$ plane -- represented by a two-parameter (single-qubit) 
gate $U_{xy}$ -- constitutes the essential single-qubit operation~\cite{Morgado+Whitlock:21}; this gate allows one to realize 
an arbitrary single-qubit rotation and -- by extension -- any single-qubit operation (e.g. the Hadamard gate).
The situation is even more favorable in the case of typical trapped-ion QC setups~\cite{Quantinuum}, where the native get 
set includes not only $x-y$ rotations but also a $ZZ$ two-qubit gate (therefore, the Ising-interaction pulse in the problem 
at hand could be implemented in that platform through a sequence of three pairwise $ZZ$ gates); in addition, the trapped-ion 
platform has the advantage of allowing perfect (all-to-all) connectivity between individual qubits.

As substantiated above, global-control- and interaction pulses required for the realization of the envisaged state interconversion
represent the basic gate operations in systems with Ising-type qubit-qubit coupling~\cite{Jones:03,Ichikawa+:13}.
For completeness, it is interesting to note that those types of pulses are formally equivalent with the two unitary operations 
utilized in the generalized form of the quantum approximate optimization algorithm (QAOA)~\cite{Farhi+:14}. More precisely, 
the Ising-interaction pulse corresponds to the cost-Hamiltonian-based unitary operator, as the Ising model encodes the cost function 
of a typical combinatorial-optimization problem (e.g. Max-Cut). At the same time, our global control pulses have the same form 
as the mixing-Hamiltonian unitary in QAOA under the assumption that the latter is generalized so as to involve not only the 
Pauli-$X$ but also $Y$ operators. The corresponding rotation angles should be the same for different qubits and only vary 
between different rounds, which is one of the already investigated modifications of the original QAOA algorithm.

\subsection{Relevant time-evolution operators}  \label{TimeEvolOps}
In what follows, we present the derivation of the time-evolution operators describing the control- and Ising-interaction
pulses enabling the $W$-to-GHZ state conversion in the three-qubit system under consideration.

Using the form of the Ising-interaction Hamiltonian [cf. Eq.~\eqref{threeQubitZZXY}] in the chosen 
symmetry-adapted basis [cf. Eq.~\eqref{BasisColumnVecs}],
\begin{equation}
H_{ZZ} \mapsto J\:\begin{pmatrix}3 & 0 & 0 & 0 \\ 0 & -1 & 0 & 0\\ 0 & 0 & -1 & 0\\ 0 & 0 & 0 & 3 
\end{pmatrix} \:, \label{HzzMatrix}\\
\end{equation}
it is straightforward to derive the time-evolution operator corresponding to the Ising interaction 
pulse (cf. Fig.~\ref{fig:PulseSequence}). This time-evolution operator is given by
\begin{equation} \label{TimeEvolutionIsing}
U_{ZZ}(\xi) = e^{-i\xi H_{ZZ}/J} \mapsto \begin{pmatrix}
e^{-3i\xi}&0&0&0\\0&e^{i\xi}&0&0\\0&0&e^{i\xi}&0\\0&0&0&e^{-3i\xi}
\end{pmatrix}\:,
\end{equation}
where $\xi\equiv JT$ is the dimensionless duration of the Ising-interaction pulse (i.e. the time 
delay between the two control pulses). 

We now address the form of the time-evolution operator of one instantaneous (delta-shaped) control pulse 
[cf. Eq.~\eqref{eq:h(t)_def}]. Even though the corresponding (time-dependent) control Hamiltonian involves 
the mutually noncommuting Pauli operators $X_n$ and $Y_n$ ($n=1,2,3$), the time-dependence of the $x$- and 
$y$ control fields is the same, which implies that this control Hamiltonian has the property of commuting 
with itself at different times (i.e. $[H_C(t),H_C(t')]=0$). Consequently, its corresponding time-evolution
operator is given by $\exp[-i\int_{t_i}^{t_f}H_C (t)dt]$ (where $t_i$ and $t_f$ are the initial and final 
evolution times, respectively), rather than requiring a time-ordered exponential (Dyson series). This 
operator is given by an 
exponential of a linear combination of the Pauli operators $X_n$ and $Y_n$ and can be evaluated using the 
well-known identity for single-qubit rotation operators
\begin{equation}\label{PauliExpIdentity}
\exp[-i\theta (\mathbf{\hat{n}}\cdot \mathbf{X})]= 
\cos\theta \mathbbm{1}_{2}- i\sin\theta\:(\mathbf{\hat{n}}\cdot \mathbf{X})  \:,
\end{equation}
where $\mathbf{X}\equiv(X, Y, Z)^{\textrm{T}}$ is the vector of Pauli operators, and $\mathbf{\hat{n}}$ is an arbitrary 
unit vector. The left-hand-side of the last equation corresponds to the rotation through an angle of $2\theta$ 
around the axis defined by the vector $\mathbf{\hat{n}}$, i.e. the rotation represented by the operator 
$R_{\mathbf{\hat{n}}}(2\theta)$.

By making use of the last identity, we obtain the time-evolution operators $U_C(\boldsymbol{\alpha})$
corresponding to individual control pulses; in the problem at hand $\boldsymbol{\alpha}=\boldsymbol{\alpha}_1$ 
for the first control pulse and $\boldsymbol{\alpha}=\boldsymbol{\alpha}_2$ for the second one. These 
time-evolution operators are of the form
\begin{equation}\label{eq:V_exponential}
U_C(\boldsymbol{\alpha})=\prod_{n=1}^3(\cos\alpha\:\mathbbm{1}_{8}
-i\sin\alpha\:\mathcal{A}_n) \:,
\end{equation}
where $\mathcal{A}_n$ ($n=1,2,3$) are auxiliary operators given by
\begin{equation}\label{operatorAn}
\mathcal{A}_n=\frac{1}{\alpha}\:(\alpha_x X_n+\alpha_y Y_n)
\end{equation}
and $\alpha\equiv \|\boldsymbol{\alpha}\|>0$ denotes the norm of the vector $\boldsymbol{\alpha}$. 
By making use of the polar coordinates in the $x$-$y$ plane, the operator $\mathcal{A}_n$
on qubit $n$ can be recast in an exceedingly simple matrix form using
\begin{equation}
\frac{1}{\alpha}\:(\alpha_x X+\alpha_y Y) = \begin{pmatrix}
0 & e^{-i\phi} \\ e^{i\phi} & 0
\end{pmatrix}\:
\end{equation}
for each qubit, where $\phi$ designates the polar angle corresponding to the vector $\boldsymbol{\alpha}$. 
By analogy with the general case represented by Eq.~\eqref{PauliExpIdentity}, an instantaneous control 
pulse in the system under consideration amounts to a global rotation through an angle of $2\alpha$ around the 
axis whose direction is specified by the unit vector $\mathbf{\hat{n}}\equiv(\cos\phi,\sin\phi,0)^{\textrm{T}}$.

To obtain a more explicit form of $U_C(\boldsymbol{\alpha})$, we perform the multiplication in 
Eq.~\eqref{eq:V_exponential} and arrive at the expression 
\begin{eqnarray} \label{exprUc}
U_C(\boldsymbol{\alpha}) &=& \cos^3\alpha\:\mathbbm{1}_8
-i\sin\alpha\cos^2\alpha\:\mathcal{S}_1\nonumber\\
&-& \sin^2\alpha\cos\alpha\:\mathcal{S}_2+i\sin^3\alpha\:\mathcal{S}_3 \:,
\end{eqnarray}
where $\mathcal{S}_1$, $\mathcal{S}_2$, and $\mathcal{S}_3$ are auxiliary operators given by
\begin{eqnarray} \label{defmathcalS}
\mathcal{S}_1 &=& \sum_{n=1}^3 \mathcal{A}_n\:, \nonumber\\
\mathcal{S}_2 &=& \sum_{n<n'} \mathcal{A}_n\mathcal{A}_{n'}\:,\\\
\mathcal{S}_3 &=& \prod_{n=1}^3\mathcal{A}_n \:.\nonumber
\end{eqnarray}
When expressed in the basis of Eq.~\eqref{BasisColumnVecs}, these operators are represented 
by the $4\times 4$ matrices
\begin{eqnarray} \label{matricsmathcalS}
P_S \mathcal{S}_1 P_S^{\dagger} &=& \begin{pmatrix}
0&\sqrt{3}\:e^{-i\phi}&0&0\\
\sqrt{3}\:e^{i\phi}&0&2e^{-i\phi}&0\\
0&2e^{i\phi}&0&\sqrt{3}\:e^{-i\phi}\\
0&0&\sqrt{3}\:e^{i\phi}&0
\end{pmatrix}\:,\notag\\
P_S \mathcal{S}_2 P_S^{\dagger} &=& \begin{pmatrix}
0&0&\sqrt{3}\:e^{-2i\phi}&0\\
0&2&0&\sqrt{3}\:e^{-2i\phi}\\
\sqrt{3}\:e^{2i\phi}&0&2&0\\
0&\sqrt{3}\:e^{2i\phi}&0&0
\end{pmatrix}\:,\notag\\
P_S \mathcal{S}_3 P_S^{\dagger} &=& \begin{pmatrix}
0&0&0&e^{-3i\phi}\\
0&0&e^{-i\phi}&0\\
0&e^{i\phi}&0&0\\
e^{3i\phi}&0&0&0
\end{pmatrix}\:,
\end{eqnarray}
where $P_S$ is the projector on the (four-dimensional) symmetric sector 
[cf. Eqs.~\eqref{SSbasis} and \eqref{BasisColumnVecs}].

The time-evolution operator $U_C(\boldsymbol{\alpha}_1)\equiv U_C(\alpha_1,\phi_1)$ corresponding to 
the first ($t=0$) control pulse and its counterpart $U_C(\boldsymbol{\alpha}_2)\equiv U_C(\alpha_2,\phi_2)$ 
that pertains to the second ($t=T$) pulse are straightforwardly obtained using Eqs.~\eqref{exprUc}-\eqref{matricsmathcalS} 
[the cumbersome -- but otherwise straightforward to derive -- final expressions are not provided here]. By 
combining the final expressions for $U_C(\alpha_1,\phi_1)$ and $U_C(\alpha_2,\phi_2)$ with the previously 
derived expression for $U_{ZZ}(\xi)$ [cf. Eq.~\eqref{TimeEvolutionIsing}], one recovers the time-evolution 
operator 
\begin{equation} \label{fullU}
\mathcal{U}(\xi,\boldsymbol{\alpha}_1,\boldsymbol{\alpha}_2)=
U_C(\alpha_2,\phi_2)\:U_{ZZ}(\xi)\:U_C(\alpha_1,\phi_1) 
\end{equation}
that corresponds to the entire pulse sequence (for an illustration, see Fig.~\ref{fig:PulseSequence}).

\section{State-conversion protocol: Results and Discussion} \label{StateConvResDisc}
In the following, we present and discuss the result for the state-conversion protocol based 
on the pulse sequence of Sec.~\ref{WtoGHZsequence}. We first discuss the results obtained 
through numerical optimization of the GHZ-state fidelity corresponding to this pulse sequence 
(Sec.~\ref{FidelityOptimize}). We then consider the robustness of the state-conversion protocol
to errors in its characteristic parameters (Sec.~\ref{PulseSequenceRobust}).
\subsection{Optimization of the target-state fidelity} \label{FidelityOptimize}
Aiming to convert the initial state $\ket{W_3}$ into $\ket{\text{GHZ}_3(\varphi)}$ for an arbitrary 
value of $\varphi$, we maximize the central figure of merit in the problem at hand -- the GHZ-state 
fidelity $\mathcal{F}_{\textrm{GHZ}}(\varphi)$ -- with respect to the parameters $\xi$, $\alpha_1$, 
$\phi_1$, $\alpha_2$, and $\phi_2$ characterizing the envisaged pulse sequence (cf. Sec.~\ref{WtoGHZsequence}). 
This fidelity is given by
\begin{equation} \label{GHZfidelity}
\mathcal{F}_{\textrm{GHZ}}(\varphi)=\left|\bra{\text{GHZ}_3(\varphi)}\,
\mathcal{U}(\xi,\boldsymbol{\alpha}_1,\boldsymbol{\alpha}_2)\ket{W_3}\right| \:,
\end{equation}
i.e. by the module of the overlap of the target state $\ket{\text{GHZ}_3(\varphi)}$ and the actual 
state $\mathcal{U}(\xi,\boldsymbol{\alpha}_1,\boldsymbol{\alpha}_2)\ket{W_3}$ obtained at the end 
of the pulse sequence [cf. Eq.~\eqref{fullU}]. Given that the target GHZ state in the state-conversion 
problem at hand is parameterized by $\varphi$ [cf. Eq.~\eqref{InitFinalStates}], it is plausible to 
expect that the values ($\xi_0$,$\alpha_{1,0}$, $\phi_{1,0}$, $\alpha_{2,0}$, $\phi_{2,0}$) of these 
parameters that correspond to the maximum of $\mathcal{F}_{\textrm{GHZ}}(\varphi)$ should also depend 
on $\varphi$.

\begin{figure}[t!]
\includegraphics[width=0.7\linewidth]{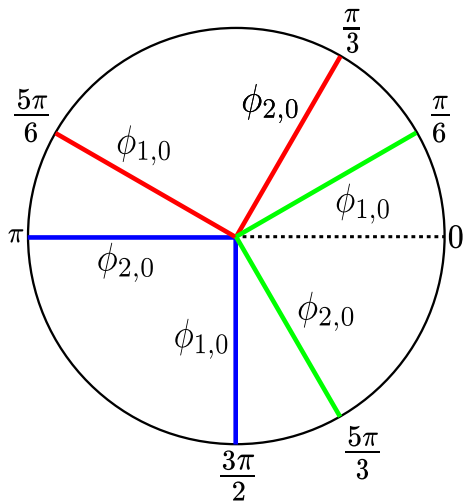}
\caption{\label{fig:ThreePhiBranches}(Color online)Pictorial illustration of the three branches of 
optimal values $\phi_{1,0}$ and $\phi_{2,0}$ of the angles that specify the directions of the 
global-rotation axes corresponding to the control pulses.}
\end{figure}

We first carry out the optimization of the fidelity in Eq.~\eqref{GHZfidelity} numerically for $\varphi=0$
using the \texttt{minimize} routine from the \texttt{scipy.optimization} package~\cite{minimize_scipy} of 
the SciPy library. In this manner, we obtain the optimal values $\xi_0=0.3077$, $\alpha_{1,0}=\pi/4$, and 
$\alpha_{2,0}=0.3077$ for the parameters $\xi$, $\alpha_1$, and $\alpha_2$, respectively. At the same time,
for $\phi_1$ and $\phi_2$ we find three different branches of optimal values
\begin{equation} \label{ThreeBranches}
(\phi_{1,0},\phi_{2,0})=\{(5\pi/6,\pi/3),(3\pi/2,\pi), (\pi/6,5\pi/3)\}\:,
\end{equation}
which correspond to three different choices for the directions of the global-rotation axes. As illustrated by 
Fig.~\ref{fig:ThreePhiBranches}, in all three cases the rotation axes corresponding to the control pulses are 
mutually perpendicular. Assuming that we choose $\phi_{1,0}=5\pi/6$ (along with $\phi_{2,0}=\pi/3$) [cf. Eq.~\eqref{ThreeBranches}], 
the first control pulse of the envisioned pulse sequence is equivalent to a global qubit rotation through an 
angle of $2\alpha_{1,0}=\pi/2$, around the axis specified by the unit vector $\mathbf{\hat{n}}_1\equiv 
(-\sqrt{3}/2,1/2,0)^{\textrm{T}}$.

The numerically obtained optimal values of $\xi_0$ and $\alpha_{2,0}$ can be made plausible in 
the following manner. By inserting the obtained values of $\alpha_{1,0}$, $\phi_{1,0}$, and 
$\phi_{2,0}$ -- along with the observation that $\xi_0=\alpha_{2,0}$ -- into the general expression 
for $\mathcal{F}_{\textrm{GHZ}}(\varphi)$ [cf. Eq.~\eqref{GHZfidelity}], we obtain
\begin{equation}
\mathcal{F}_{\textrm{GHZ}}(\varphi) = 
\frac{\sqrt{3}}{4}\sqrt{5+2\cos(4\xi_0)-3\cos^2(4\xi_0)} 
\left|\cos(\frac{\varphi}{2})\right| \:.
\end{equation}
Based on this last expression, $\mathcal{F}_{\textrm{GHZ}}(\varphi)=1$ if 
$\cos(4\xi_0)=1/3$ and $\varphi=0$. Therefore, the optimal values of $\xi$ 
and $\alpha_2$ are given by
\begin{equation}
\xi_0=\alpha_{2,0}=\frac{1}{4}\arccos \frac{1}{3} \:,
\end{equation}
which is equal to $0.3077$ found numerically.

By choosing $\phi_{2,0}=\pi/3$ (along with $\phi_{1,0}=5\pi/6$) [cf. Eq.~\eqref{ThreeBranches}] the second 
control pulse of the envisioned pulse sequence amounts to a global qubit rotation through an angle of $2\alpha_{2,0}
=\arccos(1/3)/2$, around the axis specified by the unit vector $\mathbf{\hat{n}}_2\equiv(1/2,\sqrt{3}/2,0)^{\textrm{T}}$.

Given that the envisaged pulse sequence entails two instantaneous control pulses, the total duration
of the pulse sequence is effectively given by $\xi$ -- the (dimensionless) duration of the Ising-interaction 
pulse. Therefore, to verify that the obtained value $\xi_0$ of this last parameter indeed represents
the minimal possible pulse-sequence duration that allows one to reach the GHZ-state fidelity close
to unity, we performed the following numerical check. We reduced $\xi$ to values below $\xi_0$ and tried 
to maximize $\mathcal{F}_{\textrm{GHZ}}(\varphi=0)$ with respect to the remaining four parameters. By 
so doing, we corroborated that $\xi_0=\arccos(1/3)/4$ (i.e. $T_0 = 0.3077\:J^{-1}$ upon reinstating 
the dimensionful units) is indeed the sought-after minimal pulse-sequence duration that enables one 
to carry out the desired $W$-to-GHZ state conversion.

Having obtained the optimal values of the five pulse-sequence parameters for $\varphi=0$, we performed 
numerical optimization of the GHZ-state fidelity for $100$ nonzero values of $\varphi$ in $[0,2\pi)$. 
These calculations lead to two important conclusions. Firstly, the optimal values $\alpha_{1,0},\alpha_{2,0},
\xi_0$ are completely independent of $\varphi$. Secondly, the optimal values of $\phi_{1,0}$ and $\phi_{2,0}$ 
depend linearly on $\varphi$. More specifically yet, the following linear dependencies are recovered 
from the obtained numerical results:
\begin{eqnarray}\label{phisWtoGHZ}
\phi_{1,0}&=&\varphi/3+2\pi m/3+5\pi/6\:, \notag\\
\phi_{2,0}&=&\varphi/3+2\pi m/3+\pi/3\:,
\end{eqnarray}
where $m=0,1,2$ enumerates different branches of optimal values.

From the form of the last equation it can be inferred why there are three branches of possible solutions 
for the optimal values of the parameters $\phi_1$ and $\phi_2$ [cf. Eq.~\eqref{ThreeBranches}]. Namely, 
adding multiples of $2\pi$ to $\varphi$ does not change the GHZ state itself [cf. Eq.~\eqref{InitFinalStates}], 
while it yields additional possible values of $\phi_{1,0}$ and $\phi_{2,0}$ in $[0,2\pi)$. More specifically
yet, by adding $2\pi$ to the value $\varphi=0$ (that yields $\phi_{1,0}=5\pi/6$, $\phi_{2,0}=\pi/3$) one
obtains $\phi_{1,0}=3\pi/2$, $\phi_{2,0}=\pi$, while by adding $4\pi$ one finds $\phi_{1,0}=\pi/6$, 
$\phi_{2,0}=5\pi/3$ (for an illustration, see Fig.~\ref{fig:ThreePhiBranches}). 

For the sake of completeness, having considered $W$-to-GHZ state conversion it is worthwhile to briefly 
comment on the reversed state-conversion process, i.e. the one whereby an initial ($t=0$) GHZ state 
is converted into a $W$ state ($t=T$). The first instantaneous control pulse -- acting on a GHZ state at 
$t=0$ -- is parameterized by $\alpha_2$ and $\phi_2$, while the second one (at $t=T$) is characterized by 
the parameters $\alpha_1$ and $\phi_1$. Our numerical optimization of the $W$-state fidelity -- defined by 
analogy with Eq.~\eqref{GHZfidelity} -- leads to the conclusion that the optimal values of the parameters
$\alpha_1$, $\alpha_2$, and $\xi$ remain the same, while the three branches of solutions for $\phi_{1,0}$ 
and $\phi_{2,0}$ are in this case given by
\begin{equation} \label{ThreeBranchesGHZtoW}
(\phi_{1,0},\phi_{2,0})=\{(7\pi/6, 5\pi/3), (\pi/2, \pi), (11\pi/6, \pi/3)\}\:.
\end{equation}
Finally, the counterpart of Eq.~\eqref{phisWtoGHZ} in the case of GHZ-to-$W$ state conversion reads as follows:
\begin{eqnarray}\label{phisGHZtoW}
\phi_{1,0} &=& -\varphi/3 + 2\pi m/3 + 7\pi/6 \:, \notag\\
\phi_{2,0} &=& -\varphi/3 + 2\pi m/3 + 5\pi/3\:.
\end{eqnarray}

\subsection{Robustness of the state-conversion scheme to errors} \label{PulseSequenceRobust}
Having obtained the parameter values that correspond to $W$-to-GHZ state conversion in Sec.~\ref{FidelityOptimize}, 
we now discuss the robustness of the envisaged state-conversion scheme to errors in those parameters. For definiteness, 
we mostly discuss this issue in the $\varphi=0$ case; for a generic value of $\varphi$ the discussion would be fairly 
similar.

In the NMR realm it is common to consider various imperfections in pulse-sequence realizations~\cite{Vandersypen+Chuang:05}.
They typically amount to an error in the rotation axis (i.e. the direction of its corresponding unit 
vector $\mathbf{\hat{n}}$) and/or an error in the rotation angle. Therefore, the actual qubit rotation 
applied is not the ideal $R_{\mathbf{\hat{n}}}(2\theta)\equiv\exp[-i\theta (\mathbf{\hat{n}}\cdot\mathbf{X})]$ 
[cf. Eq.~\eqref{PauliExpIdentity}], but is instead given by
\begin{equation}
\tilde{R}_{\mathbf{\hat{n}}}(2\theta)=\exp\left[-i\:\mathbf{f}(\theta,\mathbf{\hat{n}})
\cdot\mathbf{X}\right]\:,
\end{equation}
where $\mathbf{f}(\theta,\mathbf{\hat{n}})$ is a vector function that characterizes the systematic error~\cite{Vandersypen+Chuang:05}.
For instance, $\mathbf{f}(\theta,\mathbf{\hat{n}})=\theta(1+\varepsilon_{\theta})\mathbf{\hat{n}}$ describes under- and over-rotation
errors (for negative- and positive values of $\varepsilon_{\theta}$, respectively). At the same time, $\mathbf{f}(\theta,\mathbf{\hat{n}})
=\theta(n_x\cos\varepsilon_{\phi}+n_y\sin\varepsilon_{\phi}, n_y\cos\varepsilon_{\phi}-n_x\sin\varepsilon_{\phi}, n_z)^{\textrm{T}}$ 
captures an error pertaining to the direction of the rotation axis~\cite{Vandersypen+Chuang:05} whose original direction is specified 
by the unit vector $\mathbf{\hat{n}}\equiv(\cos\phi,\sin\phi,0)^{\textrm{T}}$.

In keeping with the above general considerations, it is pertinent to investigate the robustness of the $W$-to-GHZ 
state-conversion scheme based on the idealized pulse sequence described in Sec.~\ref{FidelityOptimize} to systematic 
errors. Among them, it is worthwhile to consider errors in the rotation angles corresponding to the instantaneous 
control pulses (related to the parameters $\alpha_1$ and $\alpha_2$), errors pertaining to the directions of the
attendant rotation axes ($\phi_1$ and $\phi_2$), as well as pulse-length errors of the Ising-interaction 
pulse ($\xi$). To this end, we consider errors of either sign for the five relevant parameters: 
\begin{eqnarray}
\xi &=& \xi_0(1+\varepsilon_\xi)\:, \\ 
\alpha_j &=& \alpha_{j,0}(1+\varepsilon_{\alpha_j})\:,\quad
\phi_j=\phi_{j,0}+\varepsilon_{\phi_j} \quad (\:j=1,2\:)\:.\nonumber
\end{eqnarray}
Regarding the form of the last equation, it should be noted that the introduced errors in the parameters 
$\xi$, $\alpha_1$, and $\alpha_2$ have the character of relative errors, while for $\phi_1$ and $\phi_2$ 
it is more meaningful to consider absolute errors.

For general (i.e. not necessarily small) values of $\varepsilon_\xi,\varepsilon_{\alpha_1},
\varepsilon_{\phi_1},\varepsilon_{\alpha_2},\varepsilon_{\phi_2}$, the GHZ-state fidelity 
is given by
\begin{eqnarray}\label{eq:fid_err}
\mathcal{F}_\text{GHZ}(\varphi,\varepsilon_\xi) &=& \frac{1}{4} \left| 3+e^{4i\xi_0\varepsilon_\xi} \right|, \notag\\
\mathcal{F}_\text{GHZ}(\varphi,\varepsilon_{\alpha_1}) &=& \frac{1}{2}|\cos(\alpha_{1,0}
\varepsilon_{\alpha_1})|\left|3\cos(2\alpha_{1,0}\varepsilon_{\alpha_1})-1\right|, \notag\\
\mathcal{F}_\text{GHZ}(\varphi,\varepsilon_{\phi_1}) &=& \frac{1}{8}\left|3-2e^{i\varepsilon_{\phi_1}}
+3e^{2i\varepsilon_{\phi_1}}\right|\left|1+e^{i\varepsilon_{\phi_1}}\right|, \notag\\
\mathcal{F}_\text{GHZ}(\varphi,\varepsilon_{\alpha_2}) &=& \left|\cos (2\alpha_{2,0}\varepsilon_{\alpha_2})
-\frac{i}{2}\:\sin (2\alpha_{2,0}\varepsilon_{\alpha_2})\right|,\notag\\
\mathcal{F}_\text{GHZ}(\varphi,\varepsilon_{\phi_2}) &=& \frac{\sqrt{3}}{4}\left| 
\left(1+e^{6i\varepsilon_{\phi_2}}\right)c_{-}^{3}\right.\: \notag\\
& &\qquad\left.\: + i\left(e^{i\varepsilon_{\phi_2}}+e^{5i\varepsilon_{\phi_2}}
\right) w c_+c_-^2 \right.\notag\\&&
\qquad\left. + \left(e^{2i\varepsilon_{\phi_2}}+e^{4i\varepsilon_{\phi_2}}\right) 
w c_+^2c_- \right.\notag\\&&
\qquad\left.+ 2i\:e^{3i\varepsilon_{\phi_2}} c_+^{3}\right|\:,
\end{eqnarray}
where $w$ and $c_\pm$ stand for the following constants:
\begin{eqnarray}
w=\frac{1+2\sqrt{2}i}{3}\quad,\quad c_\pm = 
\sqrt{\frac{\sqrt{3}\pm\sqrt{2}}{2\sqrt{3}}}\:.
\end{eqnarray}
It is interesting to note that none of the five expressions for $\mathcal{F}_\text{GHZ}$ in Eq.~\eqref{eq:fid_err} 
has any dependence on $\varphi$, even though the optimal values of the parameters $\phi_1$ and $\phi_2$ do depend 
on $\varphi$. This can be understood as a manifestation of the general notion that the most important global 
properties of GHZ-type states do not depend on $\varphi$~\cite{Nauth+Stojanovic:22}.

We now turn our attention to the case of small deviations ($\varepsilon_p \ll 1$) from the optimal values 
of the relevant parameters ($p=\xi,\alpha_1,\phi_1,\alpha_2,\phi_2$). By expanding the respective expressions 
for the GHZ-state fidelity in Eq.~\eqref{eq:fid_err} to the lowest nonvanishing (quadratic) order in $\varepsilon_p$, 
we obtain the following results:
\begin{eqnarray}
\mathcal{F}_\text{GHZ}(\varphi,\varepsilon_\xi) &=& 1-\frac{3}{2}\:\xi_0^2\varepsilon_\xi^2
+\mathcal{O}(\varepsilon_\xi^4) \:, \notag\\
\mathcal{F}_\text{GHZ}(\varphi,\varepsilon_{\alpha_1}) &=& 1-\frac{7}{2}\:\alpha_{1,0}^2
\varepsilon_{\alpha_1}^2 +\mathcal{O}(\varepsilon_{\alpha_1}^4) \:,\notag\\
\mathcal{F}_\text{GHZ}(\varphi,\varepsilon_{\phi_1}) &=& 1-\frac{7}{8}\:\varepsilon_{\phi_1}^2
+\mathcal{O}(\varepsilon_{\phi_1}^4) \:, \label{ExpandVareps} \\
\mathcal{F}_\text{GHZ}(\varphi,\varepsilon_{\alpha_2}) &=& 1-\frac{3}{2}\:\alpha_{2,0}^2
\varepsilon_{\alpha_2}^2 +\mathcal{O}(\varepsilon_{\alpha_2}^4) \:,\notag\\
\mathcal{F}_\text{GHZ}(\varphi,\varepsilon_{\phi_2}) &=& 1-\left(2-\frac{3}{4}\:\sqrt{6}
\right)\varepsilon_{\phi_2}^2 + \mathcal{O}(\varepsilon_{\phi_2}^4) \:.\notag\quad
\end{eqnarray}
Needless to say, the linear terms in these expansions vanish because the fidelity reaches its maximum for the considered 
values $\alpha_{1,0}=\pi/4$, $\phi_{1,0}=5\pi/6$, $\phi_{2,0}=\pi/3$, and $\xi_0=\alpha_{2,0}=\arccos(1/3)/4$ of the five 
relevant pulse-sequence parameters. Based on these values of the five relevant parameters [cf. Sec.~\ref{FidelityOptimize}], 
the prefactors of the quadratic terms in the expansions of Eq.~\eqref{ExpandVareps} can straightforwardly be determined:
$3\xi_0^2/2=0.142$, $7\alpha_{1,0}^2/2=2.159$, $7/8=0.875$, $3\alpha_{2,0}^2/2=0.142$, and $2-3\sqrt{6}/4 = 0.163$, 
respectively.

\begin{figure}[b!]
\includegraphics[width=0.995\linewidth]{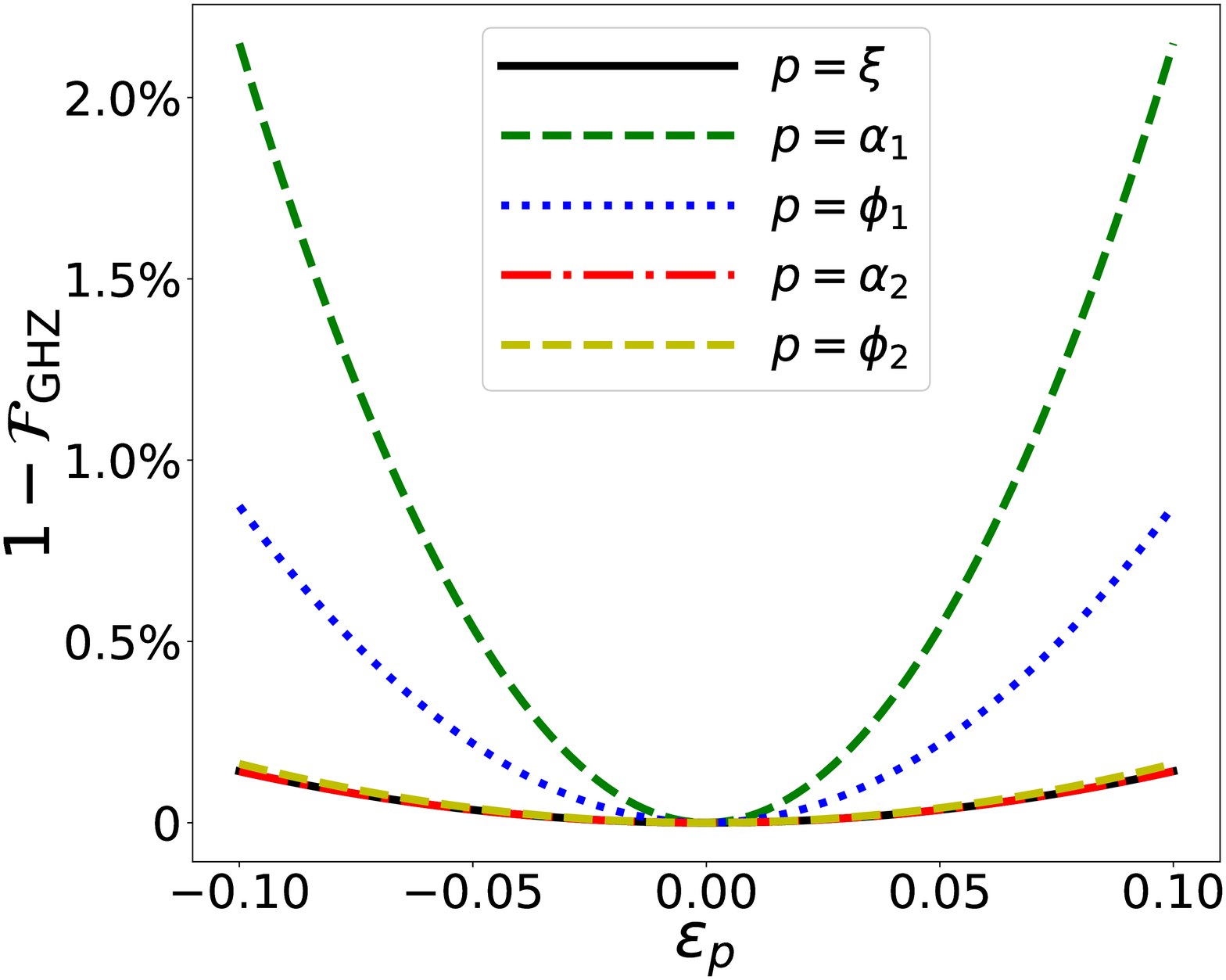}
\caption{\label{fig:InfidelityVarParams}(Color online)Deviation of the GHZ-state fidelity from unity
(i.e. the infidelity) $1-\mathcal{F}_\text{GHZ}$ as a function of errors $\varepsilon_p$ in 
the values of the parameters that characterize the pulse sequence realizing the $W$-to-GHZ state 
conversion.}
\end{figure}

The small-$\varepsilon_p$ expansions of $\mathcal{F}_\text{GHZ}$ [cf. Eq.~\eqref{ExpandVareps}], which quantify the relative 
impact on the target-state fidelity of the deviations $\varepsilon_p$ in different parameters of relevance in the problem at 
hand, are illustrated in Fig.~\ref{fig:InfidelityVarParams}. What is evident from this figure is that -- among the five 
relevant parameters -- the GHZ-state fidelity is by far most sensitive to deviations in the value of $\alpha_1$, i.e. the 
rotation angle corresponding to the first control pulse. Another salient feature of Eq.~\eqref{ExpandVareps} is that the 
obtained expansions for $\alpha_2$ and $\xi$ are completely the same (cf. Sec.~\ref{FidelityOptimize}), with the results 
for $\phi_2$ being just slightly different than those two (as can also be inferred from Fig.~\ref{fig:InfidelityVarParams}).

Another interesting conclusion can be drawn from the obtained prefactors to quadratic terms in $\varepsilon_p$ in 
Eq.~\eqref{ExpandVareps}. Namely, the respective quantitative impacts on the fidelity $\mathcal{F}_\text{GHZ}$ of errors 
in the parameters $\alpha_1$ and $\alpha_2$ (i.e. in the rotation angles corresponding to the $t=0$ and $t=T$ control 
pulses) from their optimal values differ drastically, as can also be inferred from Fig.~\ref{fig:InfidelityVarParams}. 
More precisely, for the same error (i.e. for $\varepsilon_{\alpha_1}=\varepsilon_{\alpha_2}\equiv \varepsilon_\alpha$), 
the deviation in $\alpha_1$ leads to an approximately $15$ times larger reduction of the fidelity than that of $\alpha_2$. 
Thus, our envisioned $W$-to-GHZ state-conversion scheme is much more sensitive to errors in the first control pulse 
(at $t=0$) than those corresponding to the second one (at $t=T$). This last observation can be understood
by analyzing the change in the target-state fidelity resulting from the first control pulse. Namely, this pulse 
leads to the change from $\mathcal{F}_\text{GHZ}=0$ to $\mathcal{F}_\text{GHZ}=\sqrt{3}/2\approx 0.866$, which implies
that the first control pulse represents a much bigger stride towards the final GHZ state than the second one. This makes 
the numerical results presented in Fig.~\ref{fig:InfidelityVarParams} -- i.e. the much larger sensitivity of the 
state-conversion scheme at hand to errors in the first control pulse -- completely plausible.

\begin{figure}[b!]
\includegraphics[width=0.5\textwidth]{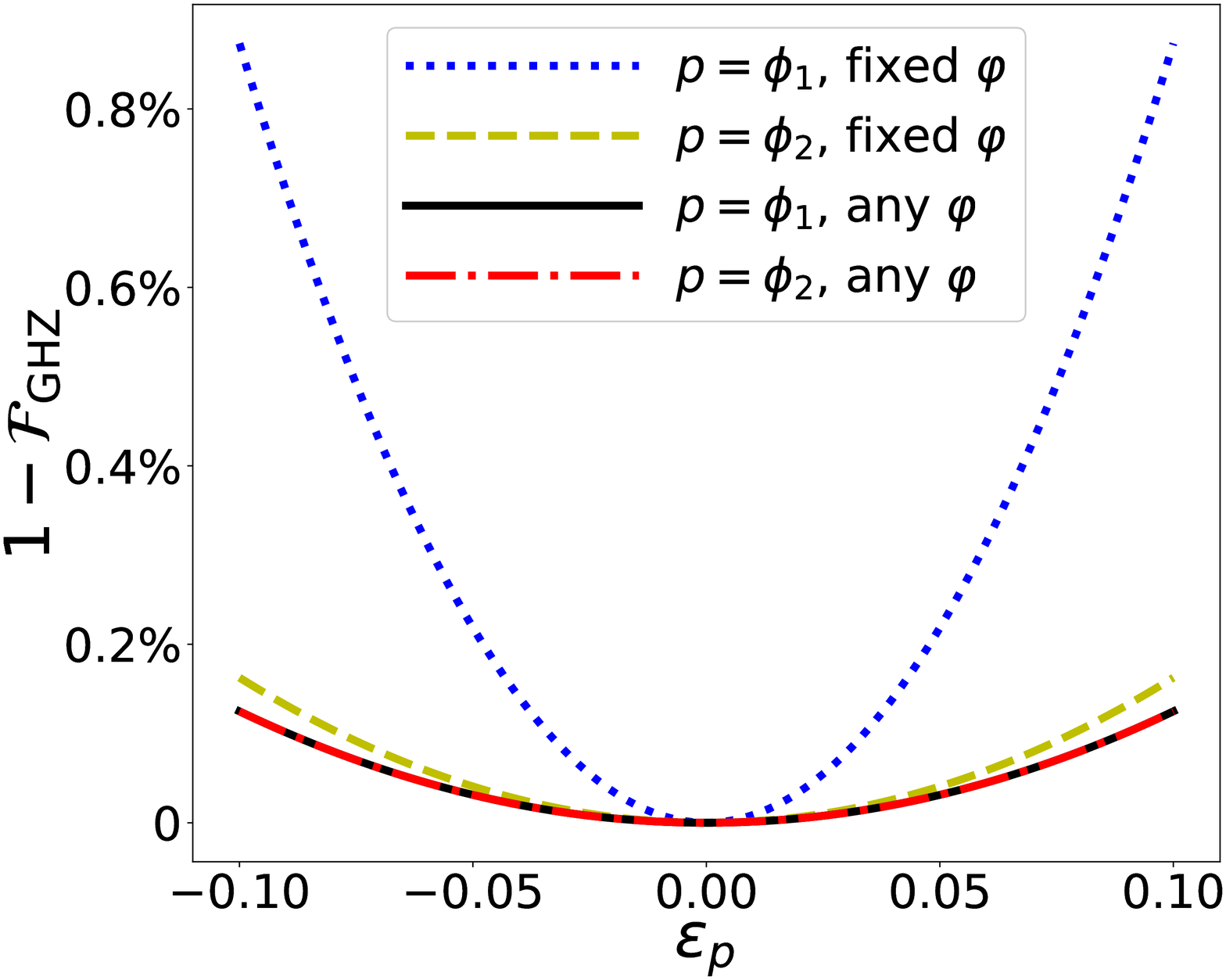}
\caption{\label{fig:InfidelityPhisArbVarphi}(Color online)Comparison of the dependence of the infidelity 
$1-\mathcal{F}_\text{GHZ}$ on $\varepsilon_{\phi_1}$ and $\varepsilon_{\phi_2}$ for fixed- and arbitrary 
value of $\varphi$.}
\end{figure}

It is well-known that the entanglement-related properties of GHZ states are largely independent of the specific 
value of $\varphi$ [cf. Eq.~\eqref{InitFinalStates}]. For instance, GHZ states are characterized by maximal essential 
three-way entanglement, as quantified by the $3$-tangle~\cite{Coffman+:00}), irrespective of the value of $\varphi$.
Likewise, these states have no pairwise entanglement, as quantified by the vanishing pairwise concurrences~\cite{HorodeckiRMP:09}. 
Because of that, it makes sense to analyze the robustness of our envisaged pulse sequence to errors in situations where 
one does not prioritize obtaining a final GHZ state with the specific value of $\varphi$, but rather a GHZ state with an 
arbitrary $\varphi$. This last scenario alleviates the impact of the errors in the parameters $\varepsilon_{\phi_1}$ and 
$\varepsilon_{\phi_2}$ -- whose optimal values depend on $\varphi$ -- on the GHZ-state fidelity in the following sense.
Namely, if only the value of one angle -- e.g., $\phi_1$ -- deviates from its optimal value $\phi_{1,0}$, the fidelity 
cannot reach unity for any $\varphi$ since the found relationship between the optimal values of $\phi_{1,0}$ and 
$\phi_{2,0}$ -- given by Eq.~\eqref{phisWtoGHZ} -- is not satisfied anymore. However, the final-state fidelity might 
increase and reach values very close to unity if $\varphi$ is allowed to vary as well. In that case, we can de-facto 
treat $\varphi$ as an additional variable parameter and try 
to optimize the final-state fidelity with respect to $\varphi$ for the fixed value of the parameter $\phi_1$ that deviates 
from its optimal value $\phi_{1,0}$. In other words, in the case of fixed $\varphi$ the fidelity is computed for a 
specific, pre-determined value of $\varphi$ and deviations from its corresponding optimal value of $\phi_1$. By contrast,
in the case that we do not prioritize obtaining a GHZ state with a specific value of $\varphi$ -- but, instead, any state 
of GHZ type -- we choose for $\varphi$ the value for which the final-state fidelity reaches its maximum, given the fixed value 
$\phi_{1,0}+\varepsilon_{\phi_1}$ of $\phi_1$ (that deviates from $\phi_{1,0}$); this last maximum, in principle, need 
not be equal to unity. Both of these scenarios are illustrated in Fig.~\ref{fig:InfidelityPhisArbVarphi}, where the relative 
impacts of the errors $\varepsilon_{\phi_1}$ and $\varepsilon_{\phi_2}$ on the fidelity $\mathcal{F}_\text{GHZ}$ in the 
aforementioned cases of fixed- and arbitrary $\varphi$ are compared. What is evident from this plot is that the deviation 
of the GHZ-state fidelity from unity due to deviations in $\varepsilon_{\phi_1}$ is drastically smaller in the latter case.

\begin{figure}[t!]
\includegraphics[width=0.995\linewidth]{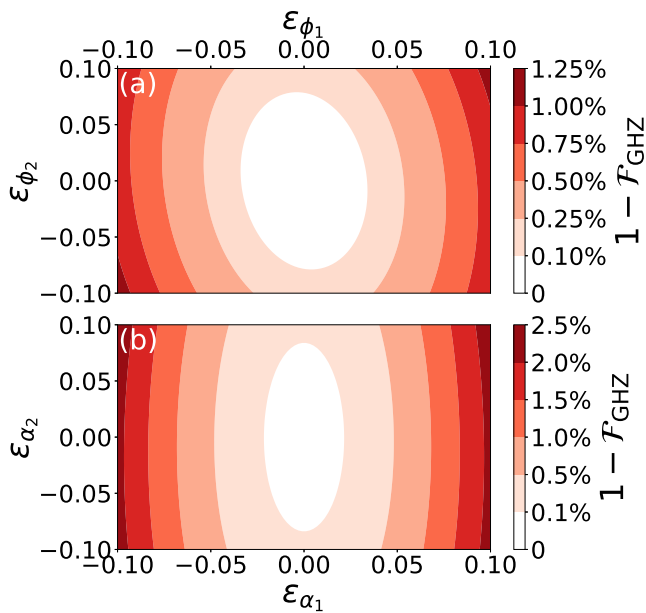}
\caption{\label{fig:InfidelityDensPlotTwoParam}(Color online)Deviation of the GHZ-state fidelity from unity
(i.e. the infidelity) $1-\mathcal{F}_\text{GHZ}$ as a function of (a) errors in the parameters $\phi_1$ 
and $\phi_2$, i.e. deviations from their respective optimal values $\phi_{1,0} = 5\pi/6$ and $\phi_{2,0}=\pi/3$, 
and (b) errors in the parameters $\alpha_1$ and $\alpha_2$, i.e. deviations from their respective optimal
values $\alpha_{1,0}=\pi/4$ and $\alpha_{2,0}=\arccos(1/3)/4$.}
\end{figure}

Aside from the expansions in Eq.~\eqref{ExpandVareps}, which quantify the impact of deviations $\varepsilon_p$ 
in individual pulse-sequence parameters on the GHZ-state fidelity, it is pertinent to also discuss the effect of 
simultaneous errors in more than one parameter. To this end, the fidelity $\mathcal{F}_\text{GHZ}$ is evaluated numerically 
based on its defining expression [cf. Eq.~\eqref{GHZfidelity}], i.e. without resorting to the small-$\varepsilon_p$ 
expansions in Eq.~\eqref{ExpandVareps}. For instance, Fig.~\ref{fig:InfidelityDensPlotTwoParam} illustrates the 
deviation $1-\mathcal{F}_\text{GHZ}$ of the fidelity from unity (the infidelity) for a range $[-0.1,0.1]$ of values for
simultaneous deviations in two different parameters. In particular, Fig.~\ref{fig:InfidelityDensPlotTwoParam}(a)
illustrates the infidelity resulting from errors in the values of the parameters $\phi_1$ and $\phi_2$, while 
Fig.~\ref{fig:InfidelityDensPlotTwoParam}(b) shows the analogous dependence on errors in the parameters 
$\alpha_1$ and $\alpha_2$. In both cases, it is noticeable that even relatively large errors (such as $0.1$)
in these parameters result in a relatively small infidelity. As can be inferred from Fig.~\ref{fig:InfidelityDensPlotTwoParam}, 
the infidelity does not exceed $1.25\%$ ($2.5\%$) in the case of parameters $\phi_1$ and $\phi_2$ ($\alpha_1$ 
and $\alpha_2$).

\begin{figure}[t!]
\includegraphics[width=0.995\linewidth]{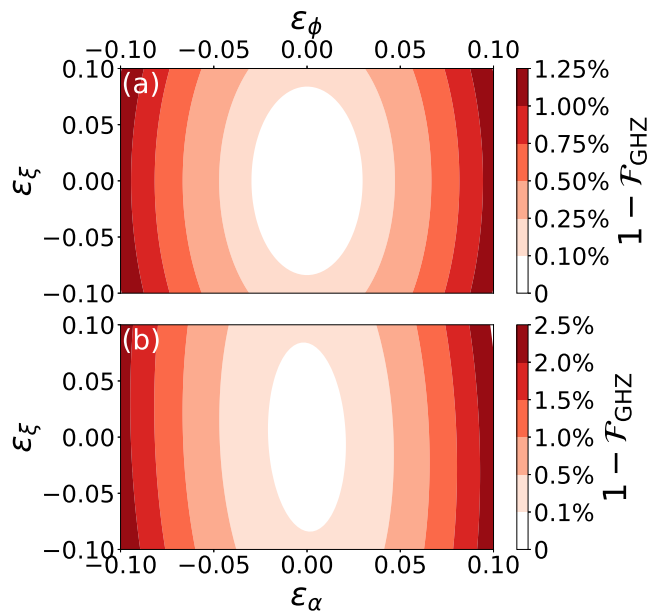}
\caption{\label{fig:InfidelityDensPlotXiPhiAlph}(Color online)Deviation of the GHZ-state fidelity from unity
(i.e. the infidelity) $1-\mathcal{F}_\text{GHZ}$ as a function of (a) errors in the Ising-pulse duration
$\varepsilon_{\xi}$ and the angles $\varepsilon_{\phi_1}=\varepsilon_{\phi_2}\equiv\varepsilon_{\phi}$ 
specifying the two rotation axes, and (b) errors in $\varepsilon_{\xi}$ and the two rotation-angles 
$\varepsilon_{\alpha_1}=\varepsilon_{\alpha_2}\equiv\varepsilon_{\alpha}$ corresponding to the  
instantaneous control pulses.}
\end{figure}

Another situation worth discussing is the one involving simultaneous errors in the Ising-pulse duration $\xi$
and the pair of parameters $\phi_1$ and $\phi_2$ (or $\alpha_1$ and $\alpha_2$). In particular, shown in 
Fig.~\ref{fig:InfidelityDensPlotXiPhiAlph}(a) is the infidelity resulting from simultaneous errors in $\xi$,
$\phi_1$, and $\phi_2$, where errors in the last two parameters are assumed to be the same (i.e. $\varepsilon_{\phi_1}
=\varepsilon_{\phi_2}\equiv\varepsilon_{\phi}$). At the same time, Fig.~\ref{fig:InfidelityDensPlotXiPhiAlph}(b) 
illustrates the infidelity resulting from errors in $\xi$, $\alpha_1$, and $\alpha_2$, where -- by analogy 
with Fig.~\ref{fig:InfidelityDensPlotXiPhiAlph}(a) -- it was assumed that $\varepsilon_{\alpha_1}=
\varepsilon_{\alpha_2}\equiv\varepsilon_{\alpha}$. What can be inferred 
from Fig.~\ref{fig:InfidelityDensPlotXiPhiAlph} is that even rather large deviations of the three relevant 
parameters -- $\xi$, $\phi_1$, $\phi_2$ in Fig.~\ref{fig:InfidelityDensPlotXiPhiAlph}(a) and $\xi$, $\alpha_1$, 
$\alpha_2$ in Fig.~\ref{fig:InfidelityDensPlotXiPhiAlph}(b) -- from their optimal values 
(up to $\varepsilon_{p}=0.1$) lead to relatively modest deviations of the GHZ-state fidelity from unity, which 
do not exceed $2.5\%$. This speaks in favor of the robustness of the envisioned $W$-to-GHZ state conversion 
to errors in the relevant parameters.

One common salient feature of Figs.~\ref{fig:InfidelityDensPlotTwoParam} and \ref{fig:InfidelityDensPlotXiPhiAlph} 
is the elliptical shape of their central, white-colored regions. This elliptical shape is a consequence of the fact 
that the lowest-order dependence of the GHZ-state fidelity on the error in the relevant parameters is quadratic. For 
instance, the lowest-order expansion of $1-\mathcal{F}_\text{GHZ}$ in $\varepsilon_{\xi}$ and $\varepsilon_{\alpha}$ 
[cf. Fig.~\ref{fig:InfidelityDensPlotXiPhiAlph}(b)] is given by
\begin{equation}
1-\mathcal{F}_\text{GHZ}(\varphi) = \frac{3}{2}\:\varepsilon_\xi^2+5\varepsilon_\alpha^2
+\varepsilon_\xi\varepsilon_\alpha + \mathcal{O}\left[(\varepsilon_\alpha,\varepsilon_\xi)^3
\right]\:,
\end{equation}
which clearly describes an ellipse in the $\varepsilon_\xi$-$\varepsilon_\alpha$ plane. The other regions 
in Figs.~\ref{fig:InfidelityDensPlotTwoParam} and \ref{fig:InfidelityDensPlotXiPhiAlph} represent dilated
versions of these central elliptically-shaped regions. 

While here we have discussed in detail the robustness to errors of the pulse sequence for converting an 
initial $W$ state into its GHZ counterpart, the robustness of the inverse (GHZ-to-$W$) state-conversion 
process can be analyzed in a completely analogous fashion.

\section{Summary and Conclusions} \label{SummConcl}
To summarize, in this paper we addressed the problem of interconversion between $W$ and GHZ states
for a three-qubit system with Ising-type coupling between qubits that are also subject to global 
transverse Zeeman-like control fields. Motivated in large part by a recent Lie-algebraic result that guarantees 
the state-to-state controllability of such a system for an arbitrary pair of initial and final states 
that are invariant with respect to permutations of qubits, we carried out our analysis within the 
four-dimensional subspace of the three-qubit Hilbert space that contains such (permutation-invariant) 
states. 

We determined a solution of the $W$-to-GHZ state-conversion problem in the form of an NMR-type pulse 
sequence, which consists of two instantaneous (global) control pulses -- each of them being equivalent 
to a global qubit rotation -- and a finite-duration Ising-interaction pulse between them. We numerically 
obtained the optimal values of the five parameters (two rotation angles corresponding to the control 
pulses, two angles that define the directions of the corresponding rotation axes, and the duration of 
the Ising-interaction pulse) that describe the envisioned pulse sequence. We then demonstrated the robustness 
of the proposed pulse sequence to errors in its five characteristic parameters. In particular, we showed 
that the GHZ-state fidelity retains values very close to unity even for appreciable deviations of the 
relevant parameters from their optimal values.

Several generalizations of the present work can be envisaged. Firstly, the robustness of the proposed 
scheme to decoherence -- an issue that necessitates treatment within the open-system scenario -- is 
worthwhile investigating. Secondly, the same deterministic interconversion problem for $W$ and GHZ states could 
also be studied for a system with more than three qubits; also, other state-interconversion problems -- involving 
various types of generalized $W$ and GHZ states, as well as other interesting classes of entangled states 
(e.g. Dicke-type states) -- are also of appreciable interest. Finally, an analogous state-interconversion problem 
could be addressed for qubit arrays with other common types of qubit-qubit interactions, such as $XY$-type 
interactions of relevance for superconducting qubits~\cite{StojanovicToffoli:12} and Heisenberg-type interactions
characteristic of spin qubits~\cite{Heule++:10,Stojanovic:19}. 

\begin{acknowledgments}
The authors acknowledge useful discussions with G. Alber. This research was supported by the Deutsche 
Forschungsgemeinschaft (DFG) -- SFB 1119 -- 236615297.
\end{acknowledgments}

\end{document}